\documentclass[fluids,article,accept,moreauthors,pdftex]{Definitions/mdpi}

\firstpage{1} 
\pubvolume{0}
\issuenum{0}
\articlenumber{0}
\pubyear{2024}
\copyrightyear{2024}
\datereceived{} 
\dateaccepted{} 
\datepublished{} 
\hreflink{https://doi.org/} 

\usepackage{amstext}
\usepackage{amsmath, float}
\usepackage{graphicx}
\usepackage{bm}
\usepackage{booktabs}

\newcommand{\varv}{v}

\newcommand{\be}{\begin{equation}}
\newcommand{\ee}{\end{equation}}
\newcommand{\bdm}{\begin{displaymath}}
\newcommand{\edm}{\end{displaymath}}
\newcommand{\bea}{\begin{eqnarray}}
\newcommand{\eea}{\end{eqnarray}}
\newcommand{\ba}{\begin{align}}
\newcommand{\ea}{\end{align}}

\newcommand       \aap          {A\&A }

\newcommand       \prd         {Phys. Rev. D}

\newcommand       \prl         {Phys. Rev. Lett.}

\Title{A GPU-Accelerated Modern Fortran Version of the ECHO Code for Relativistic Magnetohydrodynamics}

\TitleCitation{A GPU-Accelerated Modern Fortran Version of the ECHO Code for Relativistic Magnetohydrodynamics}

\Author{
  Luca Del Zanna\orcidA{} $^{1,2,3}$,
  Simone Landi\orcidB{} $^{1,2}$,  
  Lorenzo Serafini  $^{4,1}$,
  Matteo Bugli\orcidD{}  $^{5,6,7}$,  and
  Emanuele Papini\orcidC{} $^{8}$
  }

\AuthorNames{Luca Del Zanna, Lorenzo Serafini, Simone Landi, Matteo Bugli, and Emanuele Papini}

\address{%
$^{1}$ \quad Dipartimento di Fisica e Astronomia, Università degli Studi di Firenze, 50019 Sesto Fiorentino, Italy; luca.delzanna@unifi.it (L.D.Z.); simone.landi@unifi.it (S.L.)\\
$^{2}$ \quad INAF, Osservatorio Astrofisico di Arcetri, 50125 Firenze, Italy \\
$^{3}$ \quad INFN, Sezione di Firenze, 50019 Sesto Fiorentino, Italy\\
$^{4}$ \quad CINECA, 40033 Casalecchio di Reno, Italy; l.serafini@cineca.it\\
$^{5}$ \quad Dipartimento di Fisica, Università degli Studi di Torino, 10125 Torino, Italy; matteo.bugli@unito.it\\
$^{6}$ \quad INFN, Sezione di Torino, 10125 Torino, Italy\\
$^{7}$ \quad Université Paris-Saclay, Université Paris Cité, CEA, CNRS, AIM, 91191, Gif-sur-Yvette, France\\
$^{8}$ \quad INAF, Istituto di Astrofisica e Planetologia Spaziali, 00133 Roma, Italy; emanuele.papini@inaf.it
}

\abstract{
The numerical study of relativistic magnetohydrodynamics (MHD) plays a crucial role in 
high-energy astrophysics, but unfortunately is computationally demanding, given the complex
physics involved (high Lorentz factor flows, extreme magnetization, curved spacetimes near
compact objects) and the large variety of spatial scales needed to resolve turbulent motions. 
A great benefit comes from the porting of existing codes running on standard processors
to GPU-based platforms. However, this usually requires a drastic rewriting of the original code, 
the use of specific languages like CUDA, and a complex analysis of data management and 
optimization of parallel processes. Here we describe the porting 
of the \texttt{ECHO}  code for special and general relativistic MHD to accelerated devices,
simply based on native Fortran language built-in constructs, especially \texttt{do concurrent}  loops, few 
OpenACC directives, and the straightforward data management provided by the Unified Memory 
option of NVIDIA compilers.Thanks to these very minor modifications to the original code, 
the new version of \texttt{ECHO}  runs at least 16 times faster on GPU platforms compared 
to CPU-based ones. 
The chosen benchmark is the 3D propagation of a relativistic MHD Alfvén wave, 
for which strong and weak scaling tests performed on the LEONARDO pre-exascale supercomputer 
at CINECA are provided (using up to 256 nodes corresponding to 1024 GPUs, and over 14 billion cells).
Finally, an example of high-resolution relativistic MHD Alfvénic turbulence simulation is shown, 
demonstrating the potential for astrophysical plasmas of the new GPU-based version of \texttt{ECHO}.
}


\begin{document}

\section{Introduction}

The numerical investigation of high-energy astrophysical sources often requires the simultaneous
evolution of a conducting fluid and of the electric currents and magnetic fields embedded in it, typically
modeled in the regime of magnetohydrodynamics, MHD, or (general) relativistic
magnetohydrodynamics, (G)RMHD.
In particular, GRMHD simulations have recently allowed a significant advance in our understanding
of the problem of black hole accretion \cite{Porth:2019}, supporting the physical interpretation
of the Event Horizon Telescope images \cite{EHT1,EHT2}.
Given the extremely high Reynolds numbers typical of astrophysical plasmas, the regime of the fluid
is invariably turbulent, with a strongly nonlinear evolution of continuously interacting
vortices and current sheets. This scenario is for example crucial for accretion disks, where turbulence
is expected to provide both an effective viscosity/resistivity \cite{Balbus:1998,Devilliers:2003,Bugli:2018} 
and dynamo-like terms which may lead to an exponential growth of the magnetic field 
\cite{Tomei:2020,DelZanna:2022}.
This may be also important in other contexts, like neutron star merger events
\cite{Giacomazzo:2015,Ciolfi:2020}.

Direct simulations of both hydrodynamical and plasma turbulence have always been a challenging task
from a computational point of view, given the necessity of treating  several decades of the
inertial range, that is, the range of scales in which the energy
cascades from the large injection scales to those (kinetic) where dissipation takes place.
This is particularly true in relativistic MHD turbulence, since the relativistic regime adds stronger
nonlinearities to the system, due to extreme situations such as high-Lorentz factor flows and
strongly magnetized plasmas with Alfvén speed $v_\mathrm{A}$ close to the speed of light $c$.

The study of relativistic MHD turbulence is relevant in a variety of sources characterizing
high-energy astrophysics, namely accretion disks
around black holes \cite{Ripperda:2022}, high-Lorentz factor jets \cite{Mattia:2023}, 
standard and bow-shock pulsar wind nebulae \cite{Porth:2014,Olmi:2019}, to cite a few examples.
Many aspects of the relativistic MHD turbulence cascade are similar to its classic counterpart,
where the interaction of Alfvén wave packets and the deformation of eddies due to the presence
of a magnetic field, providing a natural source of anisotropy, make MHD turbulence rather
different from the hydrodynamic one.
The turbulent cascade is a channel for energy dissipation, typically occurring in very thin
current sheets where the so-called plasmoid instability may take place,
providing efficient reconnection independently of the value of actual resistivity 
\cite{Bhattacharjee:2009,Uzdensky:2010,Boldyrev:2017}.
The reconnection of thin current sheets at a fast rate independent of the Lundquist number
is also known as \textit{ideal tearing} \cite{Pucci:2014,Landi:2015,Papini:2019}, 
and it has been shown to work in the regime of relativistic MHD too \cite{DelZanna:2016}, 
providing a natural dissipation mechanism in astrophysical high-energy sources.
Another important aspect related to turbulent relativistic plasmas is that of particle acceleration,
needed to explain non-thermal emission of such sources
\cite{Guo:2014,Sironi:2014,Cerutti:2015,Beloborodov:2017,Comisso:2018,Demidem:2020,Meringolo:2023}.

Numerical studies of relativistic MHD turbulence are aimed at investigating analogies or discrepancies
with respect to classical MHD 
\cite{Zrake:2012,Takamoto:2016,Takamoto:2018,TenBarge:2021,Ripperda:2021,Chernoglazov:2021}. 
In the latter work, very high resolution 2D and 3D runs are performed in the resistive case,
in order to see the evolution of the plasmoid chains, using adaptive mesh refinement. 
In spite of this, the physical dissipation scale is still very close to the numerical one, 
and the inertial range is clearly developed in $k$-space for less than two decades.
We believe that a crucial aspect when studying turbulence is the use of a high order scheme,
whereas most of the codes, especially for relativistic MHD, reach only second order accuracy.
Higher order codes are, obviously, more computationally demanding than standard ones,
for a given number of cells; at the same time, a given level of accuracy is reached with a lower
resolution, often a crucial aspect in 3D.

Luckily, we are currently on the verge of the so-called exascale era of \textit{High Performing Computing}
(HPC), and numerical codes previously written to work on standard (CPU-based) architectures are being 
ported on the much faster (and more energetically efficient) \textit{Graphics Processing Units} (GPUs).
This is true for all kinds of research fields where fluid and plasma turbulence play a relevant role,
from engineering applications of standard hydrodynamic flows
\cite{Romero:2020,Costa:2021,Bernardini:2023,Sathyanarayana:2023,Kim:2023,DeVanna:2023}, 
to relativistic MHD and GRMHD
\cite{Grete:2021,Liska:2022,Begue:2023,Shankar:2023}.
Unfortunately, the process of porting an original code previously engineered to work on CPU-based
systems can be very long and it usually requires a complete rewriting of the code in the CUDA language
(created by NVIDIA, available for both C and Fortran HPC languages) or using meta-programming 
libraries like KOKKOS \cite[e.g.][]{Lesur:2023} or SYCL / Data Parallel C$++$
(recently successfully applied to a reduced and independent version of our code, see {{\url{https://www.intel.com/content/www/us/en/developer/articles/news/parallel-universe-magazine-issue-51-january-2023.html}}}, accessed on 27/11/2023).

Possibly slightly less computationally efficient, but offering a huge increase in portability, longevity,
and ease of programming, is to use a directive-based programming paradigm like 
OpenACC. Similarly to the OpenMP API (\textit{Application Programming Interface})
for exploiting  parallelism through multicore or multithreading programming on CPUs, 
the OpenACC API (which may also work for multicore platforms) is designed especially for GPU accelerated
devices. Specific directives, treated as simple comments by the compiler when the API is not invoked,
must be placed on top of the main computationally intensive loops, in the 
routines called inside them (if these can be executed concurrently), in the definition of arrays that will 
be stored in the device's memory, and in general in all kernels that are expected to be 
offloaded to the device, avoiding as much as possible copying data from and to CPUs.
In any case, the original code is not affected when these interfaces are not activated (via compiling 
flags). An example of a Fortran code successfully accelerated on GPUs with OpenACC
directives is \cite{DeVanna:2023} (a solver for Navier-Stokes equations).


A last possibility is the use of standard language parallelism for accelerated computing and it
would be the dream of all scientists involved in software development:
to be able to use a single version of their proprietary code, written in a standard programming language
for HPC applications (C, C$++$, Fortran), portable to any computing platform. The task of code 
parallelization and/or acceleration on multicore CPUs or GPUs (even both, in the case
of heterogeneous architectures) should be simply left to the compiler. 
An effort in this direction has been made with \textit{Standard Parallel Algorithms} 
for C$++$, but it seems that modern Fortran can really make it in
a very natural way. The NVIDIA \texttt{nvfortran} compiler fully exploits the parallelization
potential of ISO Fortran standards, for instance the \texttt{do concurrent} (DC, since 2008) 
construct for loops, that in addition to the \textit{Unified Memory} paradigm for straightforward 
data management, allows for really easy scientific coding \cite{Stulajter:2022,Caplan:2023}.
The last Intel Fortran compiler \texttt{ifx}, with the support of the OpenMP backend, 
should also allow for the offload of DC loops to Intel GPU devices.

Here we describe the rather trivial porting of the \emph{Eulerian Conservative High Order} 
(\texttt{ECHO}) code for general relativistic MHD \cite{DelZanna:2007} to accelerated devices.
We also show efficiency and scaling of the code with tests performed  on the 
LEONARDO supercomputer at CINECA, and first results of 2D relativistic MHD simulations,
for which our high order code is particularly suited.
The porting to GPUs has been mainly achieved by exploiting the standard language parallelism 
offered by ISO Fortran constructs, made possible by the capabilities of the \texttt{nvfortran} 
compiler, and by the addition of a very small number of OpenACC directives.
Given that our code has a similar structure to many other codes for fluid dynamics and MHD, 
we do hope that our positive experience will help other scientists to accelerate on GPUs, 
with a small effort, their own codes written in Fortran, the oldest programming language but,
at the same time, the most modern one, in our opinion.

The structure of this paper is the following:
section 2 will be devoted to the description of the GRMHD system of equations; in section 3
we briefly describe the code structure and explain the modifications introduced in our porting
process to GPUs; in section 4 we present and comment performance and scaling tests; 
section 5 is devoted to the first application
of the novel accelerated version of \texttt{ECHO} to relativistic MHD turbulence; section 6
contains discussion and directions for future work.

\section{General Relativistic MHD Equations in $3+1$ Form}

The GRMHD equations are a single fluid closure for relativistic plasmas in a curved spacetime,
with given metric tensor $g_{\mu\nu}$, expressing the conservation of mass and 
total (matter plus electromagnetic fields) energy and momentum
\be
\nabla_{\mu}(\rho u^{\mu}) =  0,  \qquad  
\nabla_{\mu} T^{\mu\nu} = 0,
\ee
where $\rho$ is the rest mass density, $u^\mu$ the fluid four-velocity, and $T^{\mu\nu}$
the total energy-momentum tensor.
The system also requires Maxwell's equations
\be
\nabla_{\mu}F^{\mu\nu} =  -I^{\nu}, \qquad  \nabla_{\mu}F^{*\mu\nu}  = 0,
\ee
where $F^{\mu\nu}$ is the Faraday tensor,
$F^{*\mu\nu}=\tfrac{1}{2}\epsilon^{\mu\nu\lambda\kappa}F_{\lambda\kappa}$ its dual,
$I^{\mu}$ the four-current density, 
$\epsilon^{\mu\nu\lambda\kappa}=(-g)^{-1/2}[\mu\nu\lambda\kappa]$ is the Levi-Civita pseudo-tensor
($[\mu\nu\lambda\kappa]$ is the completely antisymmetric symbol with convention $[0123]=+1$), and
$g=\mathrm{det}\{g_{\mu\nu}\}$.
Once the metric, an equation of state for the thermodynamic quantities, and an Ohm's relation for the
current are specified, the GRMHD equations are a closed system of hyperbolic, nonlinear equations.

However, for the evolution in time by means of numerical methods of the above equations, 
we need to separate time and space, abandoning the elegant and compact covariant notation.
The metric is first split in the so-called $3+1$ form, usually expressed in terms 
of a scalar \emph{lapse function} $\alpha$, a spatial \emph{shift vector} $\beta^i$, 
and the three-metric $\gamma_{ij}$, that is
\be
\mathrm{d}s^2 = \! -\alpha^2\mathrm{d}t^2+
\gamma_{ij}\,(\mathrm{d}x^i\!+\beta^i\mathrm{d}t)(\mathrm{d}x^j\!+\beta^j\mathrm{d}t),
\ee
where spatial 3D vectors (with latin indices running from 1 to 3) and tensors are those measured 
by the so-called \emph{Eulerian observer}, with unit time-like vector $n_\mu = (-\alpha, 0)$ and 
$n^\mu = (1/\alpha, -\beta^i/\alpha)$, with $\alpha=1$ and $\beta^i=0$ 
for a flat Minkowski's spacetime.
For the description of the Eulerian formalism for ideal GRMHD and the numerical implementation
in the \texttt{ECHO} code the reader is referred to \cite{DelZanna:2007}.

Using the above splitting for the metric, the (non-ideal, at this stage) GRMHD
equations in $3+1$ form for the evolution of the fluid and electromagnetic fields are
\begin{align}
 \partial_t (\sqrt{\gamma} D) & +  \partial_j [ \sqrt{\gamma} ( \alpha D v^j - D \beta^j)]  = 0, \nonumber \\
\partial_t (\sqrt{\gamma} S_i) & +  \partial_j [ \sqrt{\gamma} ( \alpha W_i^{\,j} - S_i\beta^j)]  = \sqrt{\gamma} [\tfrac{1}{2}\alpha (\partial_i\gamma_{jk}) W^{jk} +  (\partial_i\beta^j) S_j - U \partial_i\alpha], \nonumber \\ 
\partial_t (\sqrt{\gamma} U) & +  \partial_j [ \sqrt{\gamma} ( \alpha S^j  - U\beta^j)]  =  
 \sqrt{\gamma}(\alpha K_{ij} W^{ij} - S^i \partial_i \alpha), \\
\partial_t (\sqrt{\gamma} B^i) & + [ijk] \partial_j ( \alpha E_k + [klm] \sqrt{\gamma} \beta^l B^m) = 0, \nonumber \\
\partial_t (\sqrt{\gamma} E^i) & -  [ijk] \partial_j ( \alpha B_k - [klm] \sqrt{\gamma} \beta^l E^m) =  
-\sqrt{\gamma}(\alpha J^i - q\beta^i), \nonumber 
\end{align}
with the two non-evolutionary constraints
\be
\partial_i ( \sqrt{\gamma} B^i) = 0, \qquad  \,
\partial_i ( \sqrt{\gamma} E^i) = \sqrt{\gamma} q.
\ee
The above GRMHD set is then a system of 11 evolution equations for the 11 
\emph{conservative} variables $\sqrt{\gamma} [ D, S_i, U, B^i, E^i ]$, 
as measured by the Eulerian observer. 
Notice that we have let $c\to 1$ and the factor $\sqrt{4\pi}$ of Gaussian units has been absorbed 
in the electromagnetic fields, while $[ijk] $ is the 3D Levi-Civita alternating symbol.
In the above system, $D=\rho\Gamma$ is the mass density in the laboratory frame,
($\Gamma$ is the Lorentz factor of the fluid velocity $\bm{v}$), 
$\bm{S}=\rho h \Gamma^2 \bm{v} + \bm{E}\times\bm{B}$
the total momentum density, $U = \rho h \Gamma^2 -  p + \tfrac{1}{2}(E^2+B^2)$
is the total energy density,  $W_{ij} = \rho h \Gamma^2 v_iv_j
-E_iE_j - B_iB_j + p_\mathrm{tot}\gamma_{ij}$ is the total stress tensor,
$p_\mathrm{tot} = p + \tfrac{1}{2}(E^2+B^2)$, the total pressure, $\bm{E}$ and $\bm{B}$ 
are the electric and magnetic fields, whereas $q$ and $\bm{J}$ are the charge and current densities. 
Moreover, $\rho$, $p$ and $h$ are the proper mass density, pressure, and specific enthalpy.
For simplicity we assume the ideal gas law 
\be
\epsilon = \frac{p}{\hat{\gamma}-1} \Rightarrow 
h = \frac{\rho + \epsilon + p}{\rho} = 1+\frac{\hat{\gamma}}{\hat{\gamma}-1}\frac{p}{\rho},
\ee
with $\epsilon$ being the internal (thermal) energy density and $\hat{\gamma}$ the adiabatic index. 
The source term of the energy equation contains the \emph{extrinsic curvature} $K_{ij}$;
when this is not directly provided by an Einstein solver we may assume
\be
\alpha K_{ij} =  (\partial_i\beta^k)\gamma_{kj} + 
\tfrac{1}{2}(\beta^k\partial_k\gamma_{ij} - \partial_t\gamma_{ij}),
\ee
where $\partial_t\gamma_{ij} = 0$ for a stationary spacetime.
In order to close the system, an Ohm's law for $\bm{J}$ should be assigned;
see \cite{DelZanna:2022} and references therein for the implementation of the resistive-dynamo 
terms in the GRMHD equations within the \texttt{ECHO} code.

Given that the main interest here is to employ the new version of the code for applications
to wave propagation (for scaling tests) and turbulence in simple geometries
(cubic numerical domains with periodic boundary conditions),
in the following we will assume a Minkowski flat spacetime in Cartesian coordinates.
Moreover,  we impose the ideal MHD approximation, for which $\bm{E}=-\bm{\varv}\times \bm{B}$, 
so that the electric field in the fluid frame vanishes. By doing so, the electric field is no longer
an evolved quantity, and the equation for it and Gauss's law become redundant.
The system is then composed by 8 conservative equations,
that can be also written in standard 3D vectorial form as
\begin{align}
 \partial_t D & +  \bm{\nabla} \cdot ( \rho\Gamma \bm{v} ) = 0, \nonumber \\
 \partial_t \bm{S} & +  \bm{\nabla} \cdot (\rho h \Gamma^2\bm{v}\bm{v} 
-\bm{E}\bm{E} - \bm{B}\bm{B} + p_\mathrm{tot}\mathcal{I} )= 0,  \\
 \partial_t U & +  \bm{\nabla} \cdot (\rho h \Gamma^2 \bm{v} + \bm{E}\times\bm{B} )= 0, \nonumber \\
 \partial_t \bm{B} & +  \bm{\nabla} \times \bm{E} = 0, \nonumber
\end{align}
with the addition of the solenoidal constraint  $\bm{\nabla} \cdot \bm{B} = 0$.

\section{Code Structure and Acceleration Techniques}

Given that our intent is to describe the porting of a well established and tested code to GPUs, 
we will just summarize here the main features of our code, but we will not repeat validation tests, 
given that numerical methods and results are obviously not affected by the porting process.
We refer to \cite{DelZanna:2007} for the full validation of \texttt{ECHO} in both Minkowski and GR metrics.

The ideal GRMHD and RMHD equations introduced in the previous section are spatially 
discretized using finite-differences, and in semi-discrete form the evolution equations are
implemented in the \texttt{ECHO} code as
\be
\frac{d}{dt} [\mathcal{U}_i]_c =
-  \sum_j \frac{1}{h_j} ( [\mathcal{\hat{\mathcal{F}}}_i^j]_{S_j^+} - [\mathcal{\hat{\mathcal{F}}}_i^j]_{S_j^-})
+ [\mathcal{S}_i]_c, 
\ee
\be
\frac{d}{dt} [\mathcal{B}^i]_{S_i^+} =
-  \sum_{j,k} [ijk] \frac{1}{h_j} ( [\mathcal{\hat{\mathcal{E}}}_k]_{L_k^+} - [\mathcal{\hat{\mathcal{E}}}_k]_{L_k^-}), 
\ee
respectively for the equations with the divergence of fluxes and those in curl form for the magnetic field.
Here $[Q]_C$ means the point-value numerical approximation $Q(x_i,y_j,z_k)$ at cell center, for any quantity $Q$, whereas $S_j$ and $L_k$ indicate cell faces normal to direction $j$ and edges along  direction $k$, respectively.
The symbols $\pm$ indicate intermediate positions between neighbouring cells, for instance
$[Q]_{S_x^\pm} = Q(x_{i\pm 1/2},y_j,z_k)$, $[Q]_{L_z^\pm} = Q(x_{i\pm 1/2},y_{j\pm 1/2},z_k)$,
where for Cartesian grids $x_{i \pm 1/2} = \tfrac{1}{2}(x_i + x_{i \pm 1})$, and similarly for $y$ and $z$.
The peculiar discretization of the magnetic field allows to preserve the solenoidal condition 
numerically; in \texttt{ECHO} we adopt the \textit{Upwind Constrained Transport} (UCT) method. 
When UCT is not employed, the magnetic field is not staggered but it is treated as the other 
cell-centered variables, but its divergence may be non zero, especially at discontinuities where
numerical derivatives are not expected to commute.

The main points, to be computed at every substep of the iterative
Runge Kutta algorithm chosen to discretize in time the above equations, are:
\begin{enumerate}
\item If UCT is employed, the staggered magnetic fields are first interpolated at cell centers 
(INT procedure)
along the longitudinal direction (that is $B^x$ along $x$ and so on). Together with the other
conservative variables, now all defined at cell centers, we derive the primitive variables
at all locations $(x_i,y_j,z_k)$ of the numerical domain. This conservative to primitive inversion
requires an iterative Newton scheme for relativistic MHD, and it is usually the most delicate step
for all numerical codes. Both operations require a 3D loop over the entire domain.
\item For the three directions $j=1,2,3$, and for every component $i=1,\ldots 8$, the numerical fluxes 
$[\mathcal{\hat{\mathcal{F}}}_i^j]_{S_j^+}$ must be computed at all interfaces normal to $j$. 
This involves another 3D loop, for any direction. 
First we need to reconstruct primitive variables (REC procedure), then physical fluxes
are computed, later combined in second order approximations of the required
point-value numerical fluxes (they depend on the left- and right-biased reconstructed values). 
Fast characteristic speeds and contributions for UCT magnetic fluxes (electric fields) are also stored.
\item Divergence of numerical fluxes is computed, requiring another 3D loop for each direction,
and source terms are added, to complete the right hand side of equations.
Spatial derivatives along each direction are computed by calling a specific function
(DER procedure), based on neighbouring points along the considered direction.
\item If UCT is activated (UCT=YES), the upwind electric fields $[\mathcal{\hat{\mathcal{E}}}_k]_{L_k^+}$
must be computed at cell edges, using REC procedures and specific upwinding formulae.
Then their curl is computed, again invoking the DER procedure.
\item Now that all spatial derivatives have been calculated, the Runge Kutta temporal update can
be achieved, again looping on the full 3D data and for every component. Another loop is needed
to compute the timestep (actually just once per full cycle). Here the characteristic speeds previously saved
are needed, and their maximum value over the whole domain must be found.
\end{enumerate}

The numerical domain is defined by $N_x\times N_y\times N_z$ cells, to be decomposed, using the
\textit{Message Passing Interface} (MPI) library, according to the number of ranks (or tasks) for each 
direction, either selected by the user or automatically assigned by the MPI routines. 
We use an indexing different for each rank,
for instance along $x$ the cell coordinate $i$ will run from $i=1$ to $i=N_x/2$, for the first rank,
and from $i=N_x/2+1$ to $i=N_x$, for the second rank (for just two ranks along that direction).
Communication is needed before any loop containing the INT, REC, and DER routines, when
ghost cells need to be filled with data coming from a different task, along each direction.
MPI data exchange may be slow among different computation devices and, even worst, different nodes. 
However, while this is always true for REC and the computation of fluxes, 
the INT and DER procedures actually require the call to boundary conditions 
only if higher than second order accuracy is imposed (with the specific flag HO=YES).

While on CPU-based architectures the number of MPI ranks is rather free (they may correspond
to physical cores or to virtual ones), when using GPUs we found easier and more efficient to assign exactly
one MPI rank per GPU, so that communication is kept to a minimum with respect to pure computation.
In order to achieve this, a first OpenACC directive is needed.
If we have \texttt{ngpu} GPUs (and ranks) per node and \texttt{rank} is the current rank, 
after MPI initialization and after the domain decomposition, the instruction is
\\ \\
\indent !\$\texttt{acc set device\_num(mod(rank,ngpu))}
\\ \\
\noindent
and for each node, GPUs will be labeled from 0 to $\texttt{ngpu}-1$.
This instruction could actually be avoided by launching the code with a specific script,
as discussed in \cite{Caplan:2023}.

The \textit{only} other place where OpenACC directives are needed is after the declaration of
\textit{pure} routines, that is the routines which can run concurrently, called by the main 3D loops,
with no side effects among neighbouring points.
For example, the routine computing the conservative variables \texttt{u} given the
primitive variables \texttt{v} and the metric \texttt{g} (a Fortran structure, in Minkowski spacetime
it is simply a sequence of 0 or 1 values), contained in the module \texttt{physics}, is declared as 
\\ \\
\indent\texttt{pure subroutine physics\_prim2cons(v,g,u)}
\vfill !\$\texttt{acc routine}
\\ \\
\noindent
where all arguments depend on the particular point in the domain, and the computation
can be done by GPUs in any order the device will require, since at any timestep these
are independent one from another.
The same is true for the REC, DER and INT routines (actually functions in our implementation), 
acting on a one dimensional stencil of point values of a physical quantity. 
For instance, the REC function is declared as 
\\ \\
\indent\texttt{pure function holib\_rec(s) result(rec)}
\vfill !\$\texttt{acc routine}
\\ \\
\noindent
where \texttt{s} is a \textit{stencil} of neighbouring values, for example five elements for reconstruction 
with the fifth-order \textit{Monotonicity Preserving} (MP5) algorithm, 
centered at cell $i$, whereas \texttt{rec} is the reconstructed value at the $S^+$ right intercell at $i+1/2$.
This and other REC routines, as well as the INT and DER functions, are discussed in details
in the appendix of  \cite{DelZanna:2007}.
OpenACC, like OpenMP, has obviously the great advantage that a compiler not supporting those 
standards simply treats the above commands as regular comments.

Calls to such physical or interpolation-like numerical routines (or functions) must be made
from 3D loops executing solely on the GPU devices, where the order of execution is not important.
In the 2008 Fortran ISO standard, the new construct \texttt{do concurrent} (DC) was introduced,
precisely with this intent, and nowadays compilers are able to recognize it and to allow
the device (GPUs in our case) to fully exploit its accelerating potential. 
A positive experience in this direction, where standard Fortran parallelism has been implemented
in non-trivial codes for scientific use and where DC loops are basically the main upgrade
compared to a previous code's version, can be found in few other recent works 
\cite{Stulajter:2022,Caplan:2023}.

DC loops should replace all standard \texttt{do} loops and array assignements when involving
arrays defined over the whole domain pertaining to an MPI rank.  
A typical example of implementation of a DC loop in \texttt{ECHO} is
\\ \\
\indent\texttt{do concurrent (k=1:nv,iz=iz1:iz2,iy=iy1:iy2,ix=ix1:ix2)}
\vfill\hspace{5mm}\texttt{u(ix,iy,iz,k)=u0(ix,iy,iz,k)+dt*rhs(ix,iy,iz,k)}
\vfill\texttt{end do}
\\ \\
\noindent
taken from the time-stepping routine updating the array of conserved variables using the
computed right-hand-side of the equations. As we can see, a 4D DC loop is automatically collapsed
in a single one (notice that arrays are declared with direction $x$ having contiguous cells in memory,
so that multi-dimensional loops should have $x$ as the last cycling index, according to the Fortran standard
when the compiler does not recognize concurrency).
Arrays are declared and allocated by each rank as
\\ \\
\indent\texttt{real,dimension(:,:,:,:),allocatable :: u}
\nopagebreak
\vfill\texttt{allocate(u(ix1:ix2,iy1:iy2,iz1:iz2,1:nv))}
\\ \\
\noindent
where ranges \texttt{ix1}, \texttt{ix2} and the others are specific for each MPI rank
(and should include ghost cells too, for the allocation step and for most of the loops).

An important aspect to consider for effective parallelization of loops is data privatization.
It might be the case that some data are used as temporary variables private to the loop iteration:
this means that for a given a loop iteration they got a definition that is used only in that iteration.
While the compiler can automatically privatize scalar variables, it is not the case for aggregate 
data such as arrays or structures, and thus the compiler might fail to parallelize a loop by assuming 
that an array definition for a given loop iteration might be re-used in a different one.
In \texttt{ECHO} such situation occurs, for instance, when calling the REC function or the conservative
to primitive inversion routine previously mentioned.
The Fortran 202x standard adds the possibility to use the \texttt{local( )} clause to specify 
aggregate variables to privatize.
For example, the loop where numerical fluxes are computed at $x$ interfaces,
and the REC routine is called, is invoked as
\\ \\
\indent\texttt{do concurrent (iz=iz1:iz2,iy=iy1:iy2,ix=ix0:ix2) local(g,s)}
\\ \\
\noindent
where \texttt{ix0} is simply  \texttt{ix1-1} (intercells are one more with respect to cells),  
\texttt{g} is the metric structure for that particular location and  \texttt{s} is a 
1D stencil of points to be passed as input for the REC routine.
Similarly, if a same quantity inside a DC loop must be shared and updated by all independent 
threads (a \textit{reduction} operation), another special clause must be added to the DC loop.
In the \texttt{ECHO} code this is needed when computing the maximum characteristic
speed over the whole domain, and we use
\\ \\
\indent\texttt{do concurrent (iz=iz1:iz2,iy=iy1:iy2,ix=ix1:ix2) reduce(max:amax)}
\\ \\
\noindent
where \texttt{amax} is the quantity to be maximized over the parallel executed loops, 
later employed to compute the global timestep
for the Runge Kutta updates. As a warning to the reader, 
we should say that locality and reduction clauses in DC loops are recognized
by just a few Fortran compilers, and only very recently. 
The acceleration of the \texttt{ECHO} code has
also helped to improve the 2023 version of the NVIDIA \texttt{nvfortran} compiler in this respect,
thanks to S. Deldon. If other compilers are used, it is better to resort to standard
\texttt{do} loops combined with specific OpenACC directives.

The use of DC loops is of course not necessary if one prefers to employ exclusevily
OpenACC directives, in particular \texttt{!\$acc loop} combined with the \texttt{gang}, \texttt{vector}, 
and \texttt{collapse} clauses, as described for example in \cite{DeVanna:2023}. 
In the first accelerated version of our code there were no DC loops. However, since the
aim of our work was to write a version (almost) fully based on native Fortran, all \texttt{!\$acc loop} 
directives were later removed in favour of DC loops, checking that code's efficiency was not affected.
The version presented here is much simpler and even slightly faster (of a few percent factor) than
the original one based on OpenACC directives in loops.

Finally, a very powerful feature of NVIDIA compilers is the use of CUDA \textit{Unified Memory} (UM)
for dynamically allocated data. In that mode, any Fortran variable allocated in the 
standard way (dynamically, with the \texttt{allocate} statement),
will belong to a unified virtual memory space accessible by both CPUs and GPUs using same address.
When the UM mode is used, the CUDA driver is responsible for implicit data migration between 
CPU and GPU, therefore if data is used by a given device it will be automatically migrated to the 
physical memory attached to this device (not needed if data is already present).
This is not true for static data, it is thus recommended to implement any repeated heavy 
computation with DC loops using dynamically allocated variables.

In order to activate all the acceleration features on NVIDIA GPU-based platforms, we invoke
the \texttt{nvfortran} compiler with the following options
\\ \\
\indent\texttt{nvfortran -r8 -fast -stdpar=gpu -Minfo=accel}
\\ \\
\noindent
where \texttt{-r8} automatically turns all computations and declarations from \texttt{real} 
to \texttt{double precision}, \texttt{-fast} allows for aggressive optimization,
\texttt{-stdpar=gpu} enforces the Fortran standard parallelism properties on GPU devices, 
especially concurrency of DC loops, automatically offloaded on the GPUs, and
\texttt{-Minfo=accel} simply reports the success or failure of acceleration attempts.
The flag \texttt{-stdpar=gpu} is the most important one, it also automatically activates UM 
for all dynamically allocated arrays (otherwise one can invoke it with \texttt{-gpu=managed}), 
and it tells the compiler to specify OpenACC directives to GPUs (otherwise \texttt{-acc=gpu}).

\section{Performance and Scaling Results}

\begin{figure}[t]
\centering
\includegraphics[height=65mm]{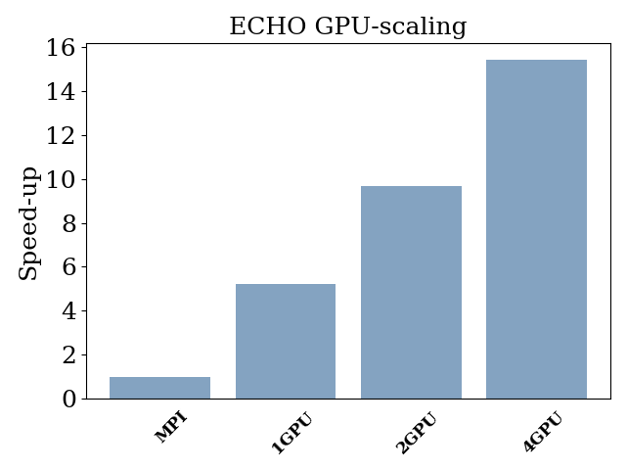}
\small
\caption{Speed-up (strong scaling) of \texttt{ECHO} for the relativistic MHD CPA-3D test
with $128^3$ cells. Using a single node of MARCONI100, we compare the time to solution
taken by 1 to 4 NVIDIA Volta V100 GPUs, compared to the time of the fastest run on the same machine
using CPUs alone (128 MPI tasks, or equivalently 32 cores with x4 hyperthreading). 
The use of GPUs over CPUs increases the speed of simulations of a factor up to nearly 16.
Results obtained by the authors in June 2022 during a NVIDIA-CINECA hackathon.}
\label{fig1}
\end{figure}

We test our accelerated version of \texttt{ECHO} on a simple but significant 3D test, the propagation
of a monochromatic, large-amplitude, circularly polarized Alfvén wave (CPA), propagating along the 
main diagonal of a periodic cubic domain of size $L$ and $N$ cells per dimension
($N^3$ in total). If $\bm{x}=(x,y,z)$ are the Cartesian coordinates for this domain, for time $t=0$ we set
the 3D-CPA wave by first defining
\be
B_X = B_0, \quad B_Y = \eta B_0 \cos(\bm{k} \cdot \bm{x}), \quad B_Z = \eta B_0 \sin(\bm{k} \cdot \bm{x}),
\ee
where $\eta$ is a non-dimensional amplitude with respect to the background magnetic field $B_0$
along direction $X$ and $k_x=k_y=k_z=2\pi/L$. The 3-velocity is set as
 $v_X=0$, $v_{Y,Z} = - v_A B_{Y,Z}/B_0$, where $v_A$ is the Alfvén speed.
In the relativistic MHD case, the value allowing for an \textit{exact} propagating CPA wave, regardless
of the amplitude $\eta$, is that reported in \cite{DelZanna:2007}, that is
\be
v_A^2 = \frac{B_0^2}{\rho h + B_0^2(1+\eta^2)}
\left[\frac{1}{2}\left(1+
\sqrt{1 - \left( \frac{2\eta B_0^2}{\rho h + B_0^2(1+\eta^2)} \right)^2} \,\,
\right)\right]^{-1},
\ee
where $\rho$, $p$ and $\rho h=\rho + 4 p$ are constant (here $\hat{\gamma}=4/3$). 
Notice that for $\eta\ll 1$ we find $v_A^2 = B_0^2/(\rho h + B_0^2)$,
and for $p\ll\rho \Rightarrow h\to 1$ and $B_0^2\ll\rho$ we retrieve the Newtonian limit of classical MHD 
$v_A^2=B_0^2/\rho$.
For our test we chose $\rho=p=B_0=\eta=1$, so that the wave has a really large amplitude and the
equations are tested in the fully nonlinear regime. In order to trigger all components and
to test the code in a truly 3D situation, the velocity and magnetic field vectors are rotated
so that $X$ is the main diagonal of the $(x,y,z)$ cubic domain, as explained in \cite{Mignone:2010}.
A full period is $T=2\pi/(kv_A)$, where $L=1$ hence $k=2\pi\sqrt{3}$ in our case.
The present set-up is probably the best possible test to check the accuracy of a classic or
relativistic MHD code, given that all variables are involved. During propagation along $X$,
the transverse components change by preserving their shape and amplitude (leaving density,
pressure, $v^2$ and $B^2$ constant), and precisely after one period $t=T$
the situation is expected to be identical to that for $t=0$. The fifth order of spatial accuracy
of \texttt{ECHO}, attainable when employing REC with MP5, with HO and DER activated, 
has been already demonstrated in \cite{DelZanna:2007} and it will not be repeated.

The authors of the present paper had the chance to participate to a three-day \textit{hackathon} 
organized (online) by the Italian supercomputing centre CINECA and by NVIDIA, in June 2022.
This was the very first time the modern Fortran version of the code was tested on GPUs, specifically 
the Volta V100 accelerators, 4 GPUs per node, of MARCONI100 (now dismissed), that also had 32 
IBM POWER9 cores  per node (with x4 hyperthreading). On that occasion, 
we successfully managed to attain the performances reported in Fig.~\ref{fig1}, where the speed-up of the
code using GPUs (with one MPI task per GPU, as explained previously) rather than just CPUs is plotted.
The numbers refer to the time to solution (one period $t=T$) for the RMHD CPA-3D test 
(here for a small resolution corresponding to $128^3$ cells) for the
case with 128 MPI tasks on CPUs (basically equivalent to 32 tasks and hyperthreading activated 
by means of \texttt{nvfortran} compiling options), divided by the time using $n$ GPUs (and $n$ tasks),
with $n=1, 2, 4$. As we can see, on a single node we achieve a x5 speed-up using 1 GPU
with respect to the best possible result using CPUs alone, and
a nearly $\times 16$ factor using all 4 GPUs.
The scaling for $n=1, 2, 4$ is rather satisfying and this can be improved by enhancing $N$ 
(the fraction of time spent in MPI communication over time spent in computation becomes lower).

It is actually difficult to make a fair comparison between GPUs and CPUs, given that on the
same machine the use of the same compiler may give non-optimal results.
Hence, we also compared our best result using 4 NVIDIA GPUs on one node (of LEONARDO, see below)
against the 32 Intel Xeon Platinum (2.4 GHz) cores of a single node of GALILEO100, using 
the appropriate Intel compiler with interprocedural and aggressive optimizations (\texttt{ifort -O3 -ip -ipo}).
The resulting speed-up was very similar, a factor $\sim 15-16$, again in favour of GPUs.
Different speed-up results can be obtained at different resolutions. 
If we compare GPU and CPU-based runs on LEONARDO, using the \texttt{nvfortran} compiler
in both cases, the speed-up is even higher: $19.2$ for a $128^3$ run, $28.6$ for a $256^3$ run, and
$30.3$ for a $512^3$ run.
Notice that the above numbers refer to runs where data output, UCT, and HO options
are not activated, so that MPI calls are just required before the main loops with REC routines
needed to compute upwind fluxes.

\begin{figure}[t] 
\centering
\includegraphics[height=48mm]{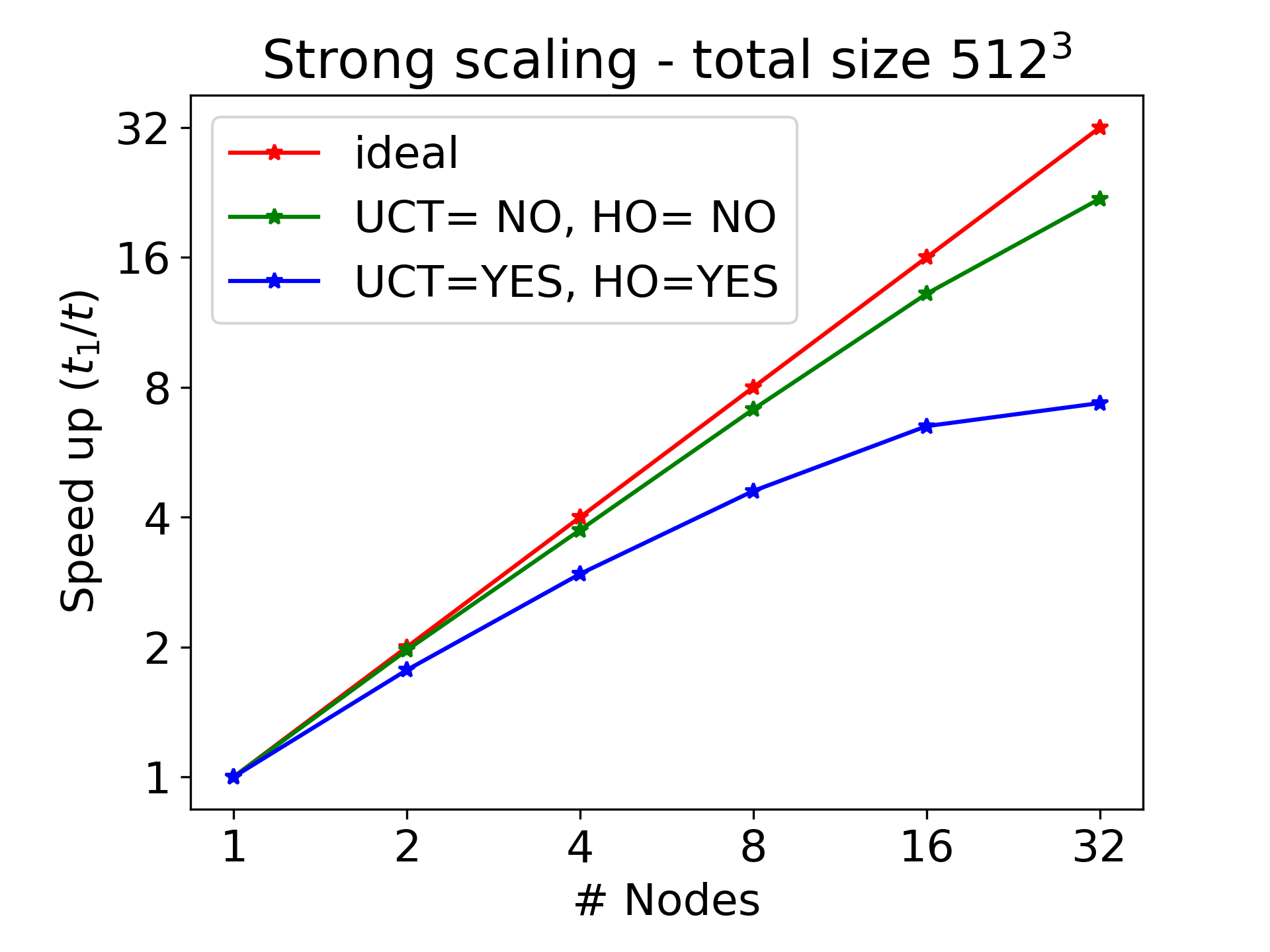}\quad
\includegraphics[height=48mm]{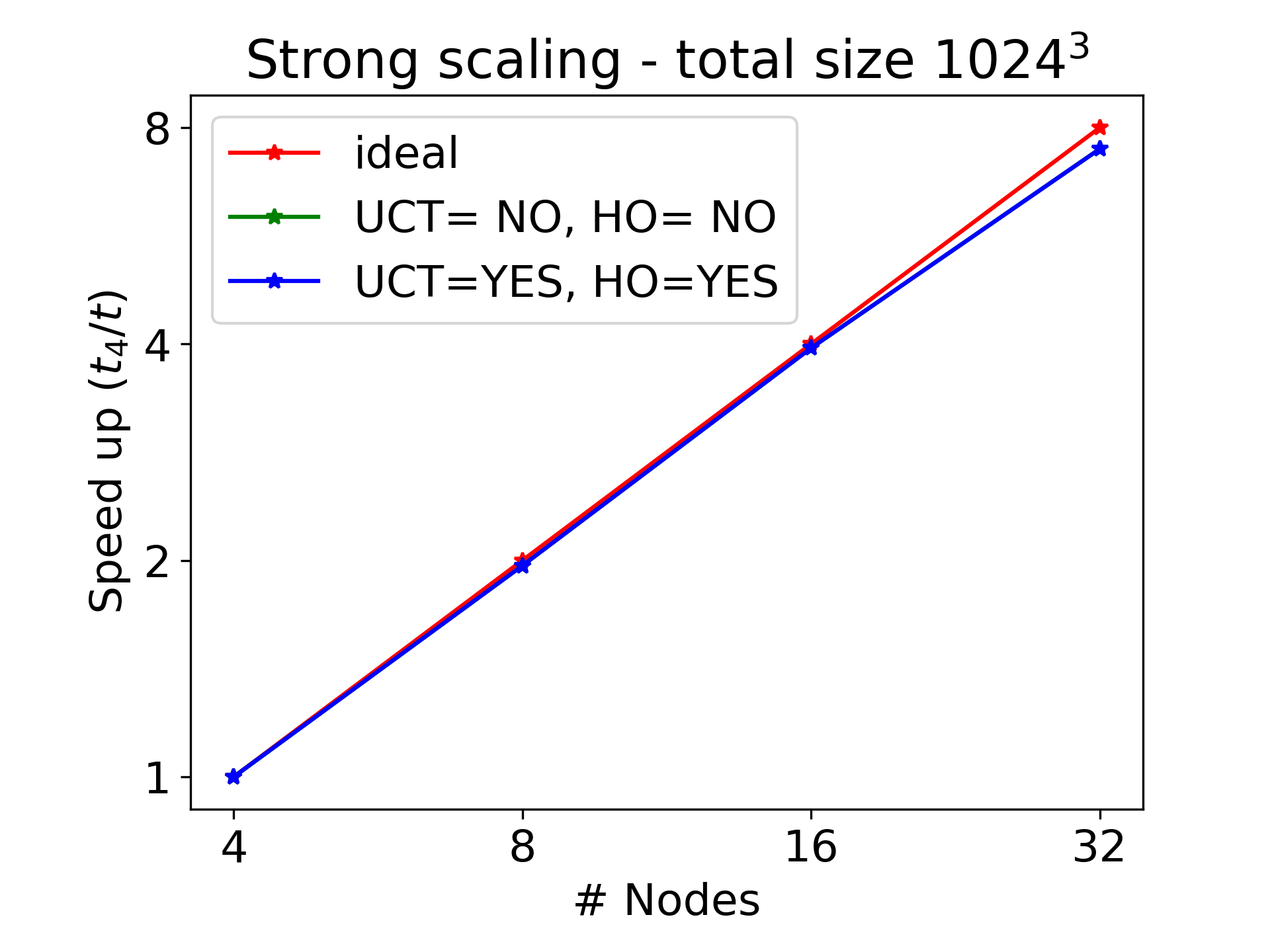}
\small
\caption{Strong scaling for the RMHD CPA-3D wave test, on LEONARDO at CINECA (4 NVIDIA
Ampere A100 GPUs per node). 
Runs with $512^3$ ($1024^3$) are displayed on the left (right) panel.
In red we show the ideal scaling, in green the speed-up in the case UCT=HO=NO 
(undistinguishable from the ideal case at $1024^3$), in blue the speed-up in the case UCT=HO=YES.}
\label{fig2}
\end{figure}

As far as scaling tests for different numbers of GPUs are concerned, we use the pre-exascale Tier-0
LEONARDO supercomputer at CINECA, equipped with 4 NVIDIA Ampere A100 GPUs and 
64 GB of memory in every node, connected by a 2000 Gb/s internal network.
We test the RMHD CPA-3D wave, now with a much larger number of cells ($512^3$ and $1024^3$),
without writing any output on disk and in two cases, with UCT and HO activated, or not.
Results are reported in Tables~\ref{tab1} and \ref{tab2}, whereas the corresponding strong scaling plots 
(time to solution with the minimun number of nodes, divided by the time to solution 
for a given number of nodes) can be found in Fig.~\ref{fig2}. 
Notice that a single node is capable of treating runs with $512^3$ cells, whereas when using $1024^3$
(more than one billion) cells, at least 4 nodes are needed, given that the memory per node is
not sufficient.

\begin{specialtable}[H] 
\small
\centering
\begin{tabular}{ccccc}
\toprule
\textbf{CPA-3D (512$^3$)} &  \textbf{UCT=HO=NO}	&	  & \textbf{UCT=HO=YES}	&  \\
\midrule
\textbf{\# Nodes}  &   \textbf{W-time [s]}  & \textbf{Speed [iter/s]}  &  \textbf{W-time [s]}  &  \textbf{Speed [iter/s]} \\
\midrule
  1		&     1735.5		& 	 1.21 	& 	1857.4		& 	 1.13  \\
  2		&  	 882.4		& 	 2.38 	& 	1050.0		& 	 2.00  \\
  4		&       464.6		& 	 4.52 	& 	  628.7		& 	 3.34  \\
  8		&      	 243.9		& 	 8.61 	& 	  403.8		& 	 5.20  \\
16		&      	 131.7		&     15.95 	& 	  285.7		& 	 7.35  \\
32		&  	   79.3		&     26.49 	& 	  252.6		& 	 8.31  \\
\bottomrule
\end{tabular}
\caption{Results for the RMHD CPA-3D wave test (using $512^3$ cells), on LEONARDO at CINECA
(4 NVIDIA Ampere A100 GPUs per node, so from 4 to 128 GPUs). 
We report all-clock time (time to solution) in seconds and speed in iterations per second 
for increasing number of nodes.}
\label{tab1}
\end{specialtable}

\begin{specialtable}[H] 
\small
\centering
\begin{tabular}{ccccc}
\toprule
\textbf{CPA-3D (1024$^3$)} &  \textbf{UCT=HO=NO}	&	  & \textbf{UCT=HO=YES}	&  \\
\midrule
\textbf{\# Nodes}  &  \textbf{W-time [s]}  & \textbf{Speed [iter/s]}  &  \textbf{W-time [s]}  &  \textbf{Speed [iter/s]} \\
\midrule
  4		& 	 7188.5		& 	 0.58 	& 	  7964.6		& 	 0.53  \\
  8		& 	 3654.7		& 	 1.16 	& 	  4212.0		& 	 1.00  \\
16		&	 1820.5		&       2.31 	& 	  2297.1		& 	 1.83  \\
32		&    	   962.0		&       4.37 	& 	  1374.4		& 	 3.06  \\
\bottomrule
\end{tabular}
\caption{Results for the RMHD CPA-3D wave test (using $1024^3$ cells), on LEONARDO at CINECA
(4 NVIDIA Ampere A100 GPUs per node, so from 16 to 128 GPUs). 
We report all-clock time (time to solution) in seconds and speed in iterations per second 
for increasing number of nodes.}
\label{tab2}
\end{specialtable}

As we can see, strong scaling is quite good when UCT and HO are not active, and the reason is that
MPI communications are kept at a minimum, only needed before the main loops for computing fluxes.
On the other hand, the UCT for preserving a solenoidal magnetic field and the use of the 
high order version of INT and DER routines require many other MPI calls to fill in ghost cells, 
and the time spent in communications over pure computation can affect efficiency severely.
This is confirmed by the fact that the problem is less important when increasing $N$,
as clearly shown in Fig.~\ref{fig2} (for $N=1024$ there is an excellent scaling even when UCT=HO=YES).
In order to alleviate the problem one should try to reduce communication at the price of increasing computation.
One way to achieve this would be to double the size of the ghost cells regions to be exchanged
between MPI tasks (before the main loops with the calls to REC), hence computing numerical fluxes
on a larger domain, so that the higher order corrections do not require any additional
update of ghost cells via MPI calls (this is true also for the UCT procedures).
This procedure would be needed just once per iteration sub-step, and would be applied to all
variables at once.

\begin{specialtable}[b] 
\small
\centering
\begin{tabular}{ccccc}
\toprule
\textbf{CPA-3D (512$^3/$node)} 	&  \textbf{UCT=HO=NO}	&	  & \textbf{UCT=HO=YES}	&  \\
\midrule
\textbf{\# Nodes}  &  \textbf{W-time [s]}  & \textbf{Speed [iter/s]}  &  \textbf{W-time [s]}  &  \textbf{Speed [iter/s]} \\
\midrule
   1		 & 	1735.5		& 	 1.21 	& 	  1857.4		& 	 1.13  \\
   2		 &  	1777.3		& 	 1.18 	& 	  1979.1		& 	 1.06  \\
   4		 & 	1804.0		& 	 1.16 	& 	  2065.5		& 	 1.02  \\
   8		 &   	1839.7		& 	 1.14 	& 	  2124.0		& 	 0.99  \\
 16		 &   	1846.8 		&       1.14 	& 	  2108.4		& 	 1.00  \\
 32		 &  	1858.7		&       1.13 	& 	  2157.2		& 	 0.97  \\
 64		 &  	1886.8		&       1.11 	& 	  2312.9		& 	 0.91  \\
128		 &  	1962.0		&       1.07 	& 	  2488.3		& 	 0.84  \\
256		 &  	1992.8		&       1.05 	& 	  2668.2		& 	 0.79  \\
\bottomrule
\end{tabular}
\caption{Results for the RMHD CPA-3D wave test (using $512^3$ cells \textit{per node}), 
on LEONARDO at CINECA (4 NVIDIA Ampere A100 GPUs per node, so from 4 to 1024 GPUs). 
We report wall-clock time (time to solution) in seconds and speed in iterations per second 
for increasing number of nodes.}
\label{tab3}
\end{specialtable}

\begin{figure}[t]
\centering
\includegraphics[height=65mm]{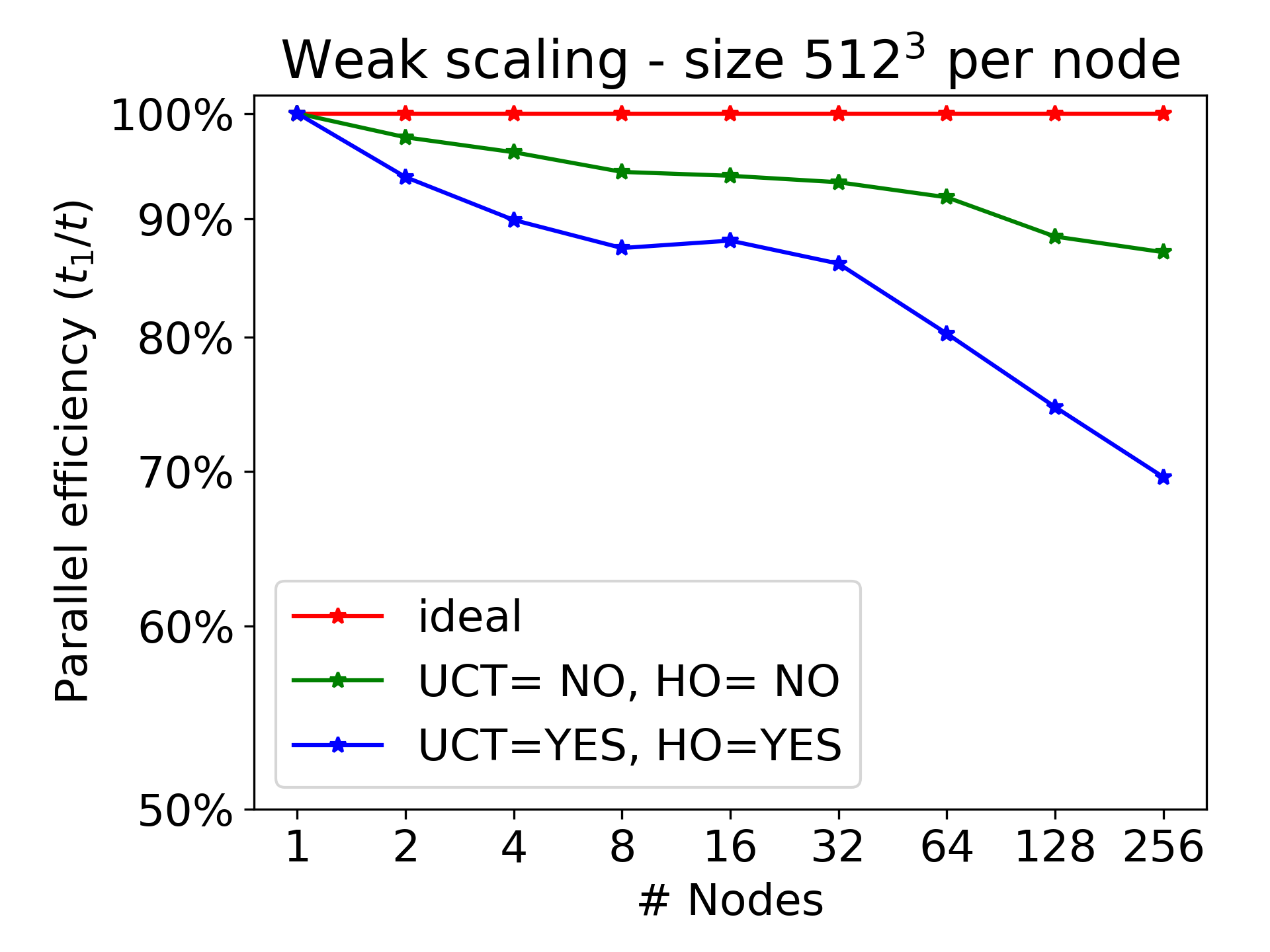}
\small
\caption{Weak scaling for the relativistic MHD CPA-3D wave test, on LEONARDO at CINECA (4 NVIDIA
Ampere A100 GPUs per node, so from 4 to 1024 GPUs). 
Parallel efficiency is defined as the inverse of the time to solution for $n$ nodes, 
normalized against the time to solution for one node, 
when each node updates a constant number of cells, here $512^3$ 
(each task, and hence each GPU, updates about 4.2 million cells).
In red we show the ideal scaling, in green the efficiency in the case UCT=HO=NO,
in blue the efficiency in the case UCT=HO=YES.}
\label{fig3}
\end{figure}

Let us now discuss weak scaling, that is the time to solution as a function of the number of nodes
(we recall that we always employ all four GPUs available per node), normalized to the case of a single node, 
when the number of computational cells per node is kept constant (here we choose $512^3$).
Results are reported in Table~\ref{tab3} and displayed in Fig.~\ref{fig3}, ranging from 1 to 256 nodes,
the maximum allowed for production runs on LEONARDO.
The largest run employing 256 nodes (hence 1024 GPUs) corresponds to
$4096 \times 4096 \times 2048  \simeq 34.4$ billion computational cells.

Since when changing the resolution also the timestep changes, for the present efficiency test
we must maintain the number of iterations constant (we choose the one needed to complete the test
on a single node).
As for strong scaling, efficiency is very good in the case UCT=HO=NO, with losses kept within $10\%$,
while it becomes much worst when communications are more numerous, in the case UCT=HO=YES.
This is just the same problem encountered for strong scaling, as discussed previously 
it could be cured by passing all boundary cells at once, at the price of increasing the computational 
domain for each node.
When this will be achieved, the corresponding efficiency is expected to stay even above 
the actual green curve in the plot.

In absolute numbers, the peak performance for the test with $512^3$
cells corresponds to $1.6 \times 10^8$ cells updated each iteration per second
(one node, UCT=HO=NO).
If we use second order time-stepping and linear reconstruction
rather than MP5, this number rises up to $2.2 \times 10^8$.
Very similar values are found by increasing even more the number of cells, up to the extreme
resolution of $640^3$ per node.

\section{A physical application: 2D relativistic MHD turbulence}

A perfect physical application of our accelerated version of the \texttt{ECHO} code is that of
decaying RMHD turbulence, given that the Minkowski flat spacetime and periodic boundary conditions
are involved, that is the same numerical setup used for our scaling benchmarks with the 
CPA-3D wave. Here only a high-resolution, 2D problem will be presented,
with fluctuations in the $x-y$ plane and a dominant, uniform magnetic field $B_0$ in the 
perpendicular direction $z$ (it is actually a 2.5D problem).
The initial condition is that of a static and homogeneous fluid with $\rho_0=1$, whereas 
(cold) magnetization and plasma-\textit{beta} parameters (as defined in \cite{DelZanna:2016}) are
\be
\sigma_0 = B_0^2/\rho_0 = 100, \quad \beta_0 = 2p_0/B_0^2 = 1,
\ee
so that $B_0=10, p_0=50$. The (relativistic) Alfvén and sound speeds are similar
\be
v_A = \frac{B_0}{\sqrt{\rho_0 + 4 p_0 + B_0^2}} \simeq 0.57, \quad
c_s = \sqrt{\frac{4}{3}\frac{p_0}{\rho_0 + 4 p_0}} \to \frac{1}{\sqrt{3}} \simeq 0.58,
\ee
where we have used $\hat{\gamma}=4/3$, corresponding to a magnetically dominated
and extremely hot plasma.
To this uniform background we add fluctuations in the perpendicular $x-y$ plane in the form
of a superposition of modes at different wavelengths
\be
\delta\bm{v} = \,\,\, \eta v_A \sum_{k_x=0}^{4} \sum_{k_y=-4}^{4} \frac{\bm{k}}{k}
\cos [\bm{k}\cdot\bm{x} + \varphi_v(\bm{k})] \, \times \bm{e}_z, 
\ee
\be
\delta\bm{B} = \eta B_0 \sum_{k_x=0}^{4} \sum_{k_y=-4}^{4} \frac{\bm{k}}{k}
\cos [\bm{k}\cdot\bm{x} + \varphi_B(\bm{k})] \, \times \bm{e}_z.
\ee
Here $\eta$ is a normalized amplitude, $\bm{k}=(k_x,k_y)$ is the wave vector
of module $k$, that for a square domain with $L_x=L_y=L=2\pi$ has components
ranging from 1 (full wavelength) to 4, whereas $\varphi_v(\bm{k})$ and $\varphi_B(\bm{k})$
and random phases depending on $k_x$ and $k_y$, different for $\bm{v}$ and $\bm{B}$.
Fluctuations are chosen to be Alfvénic, to satisfy 
$\nabla\cdot\bf{v}=\nabla\cdot\bf{B}=0$, and to have a flat isotropized spectral
radius of 4.
The common normalized amplitude has been computed (\textit{a posteriori}) 
by imposing that 
\be
\frac{< | \delta \bm{v} | >}{v_A} = \frac{< | \delta \bm{B} | >}{B_0} = \eta_\mathrm{rms} = 0.25,
\ee
where spatial averages refer to the whole 2D domain.
For the present run we employ a single node (4 GPUs) of LEONARDO at CINECA
and a resolution corresponding to $4096^2$ cells. 
The wall-clock time is about 100 seconds for simulation time unit ($t=1$ is the light crossing time
corresponding to a unit length $l=L/2\pi=1$),
using the most accurate (but also slower) version of the code with UCT=HO=YES.

\begin{figure}[b] 
\centering
\includegraphics[height=50mm]{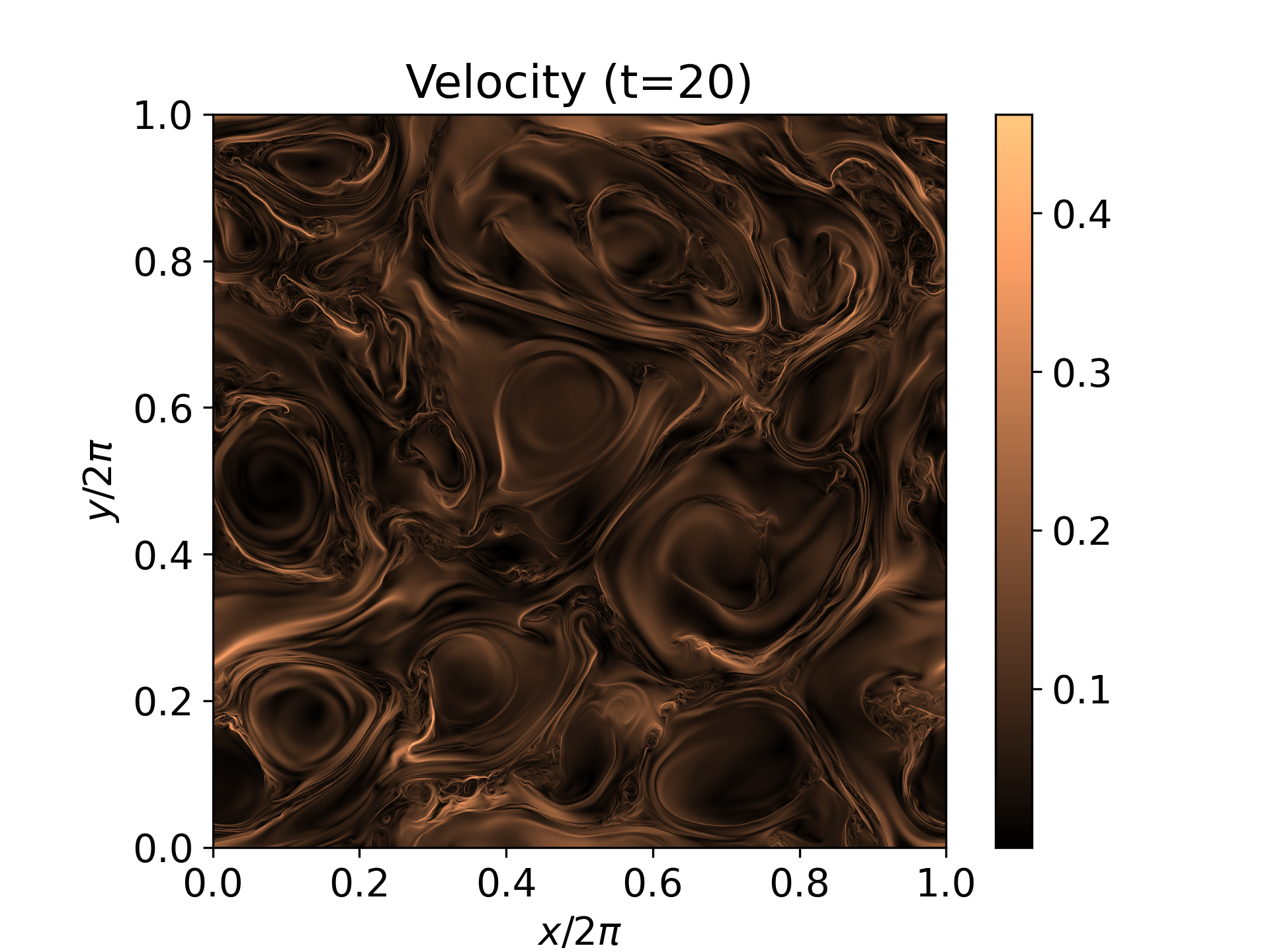}
\includegraphics[height=50mm]{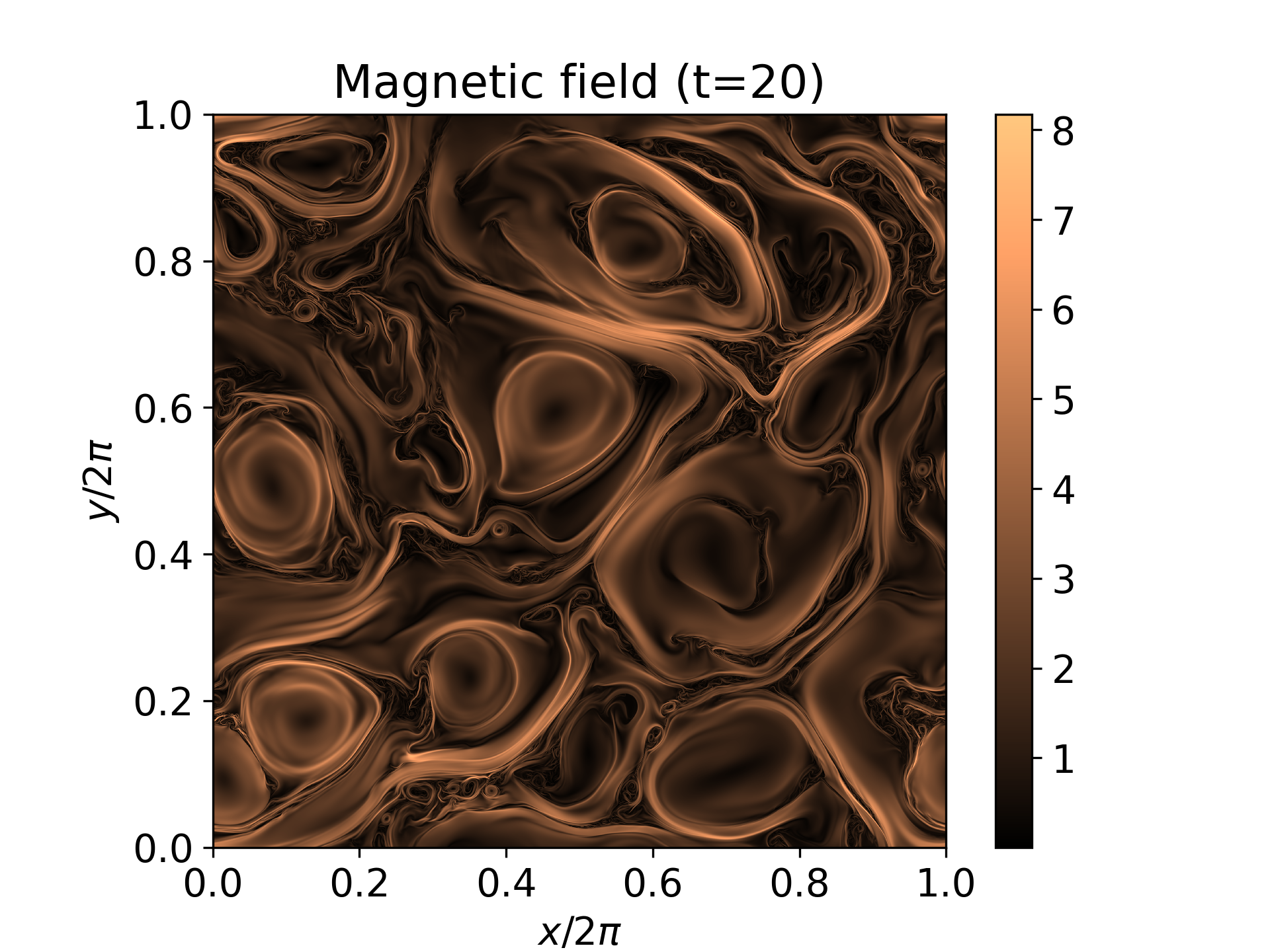}
\small
\caption{Strength of velocity and magnetic fluctuations in the perpendicular $x-y$ plane 
for a $4096^2$ run of Alfvénic 2D-RMHD turbulence, at $t=20$. Parameters and normalizations
are discussed in the text. A single node (4 GPUs) of LEONARDO at CINECA has been employed.}
\label{fig4}
\end{figure}

The simulation is run until $t=20$ (half an hour of wall-clock time), corresponding to
few nonlinear times (i.e. the characteristic evolution time of a turbulent plasma),
roughly at the peak of turbulent activity, that is when the curve of the root-mean square of
fluctuations against time reaches a maximum.
The strength of fluctuations for both the velocity and magnetic field vectors are shown 
in Fig~\ref{fig4}, and the turbulent nature of these fields is apparent. 
Vortices are now present at all scales, and energy is decaying from the large injection scale
($k\leq 4$) to the dissipative one, here simply provided by the finite precision of the 
numerical scheme, since there are no explicit velocity or magnetic dissipation terms 
in the RMHD equations solved here.
The turbulence is sub-sonic and sub-Alfvénic, so strong shocks are not present at injection scales,
even if these forms at small scales in reconnecting regions (current sheets).

\begin{figure}[t] 
\centering
\includegraphics[height=50mm]{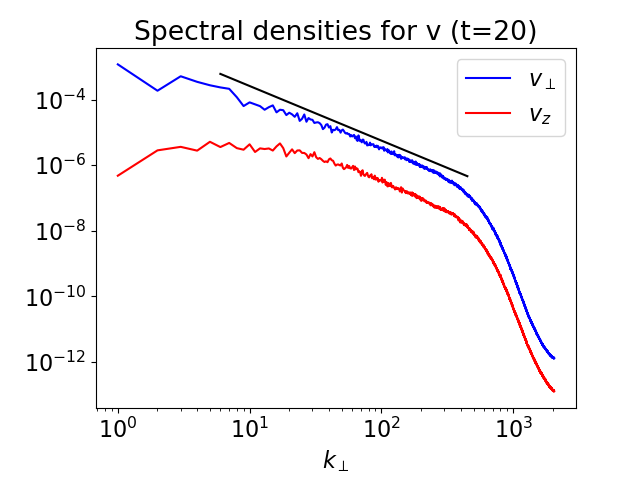}
\includegraphics[height=50mm]{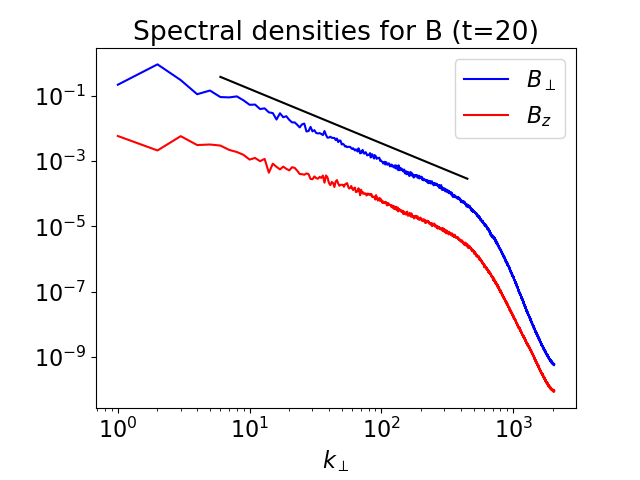}
\small
\caption{Spectral densities of velocity and magnetic fluctuations of the perpendicular ($x-y$)
and parallel ($z$) components for a $4096^2$ run of Alfvénic 2D-RMHD turbulence, at $t=20$. 
The black line refers to the Kolmogorv slope $-5/3$. Parameters and normalizations
are discussed in the text. A single node (4 GPUs) of LEONARDO at CINECA has been employed.}
\label{fig5}
\end{figure}

In Fig~\ref{fig5} we show the spectral densities for velocity and magnetic field fluctuations
as a function of $k_\perp = (k_x^2+k_y^2)^{1/2}$ (1D isotropized spectra).
In spite of the lack of physical dissipation, there is no apparent accumulation of
energy at the smallest scales, hence the numerical dissipation is rather small here.
The spectra of the perpendicular components (the dominant ones)
reproduce very well the Kolmogorov spectral slope of $-5/3$, for a couples of decades
in $k_\perp$, which is a very encouraging result for a $4096^2$ simulation with uniform grid.
We believe that the small numerical dissipation scales and the extended spectral inertial
ranges are due to the use of a scheme with high (fifth-order) spatial accuracy, one
of the main features of the \texttt{ECHO} code.

\section{Conclusions}

In the present paper we have demonstrated how a state-of-the-art numerical code
for relativistic magnetohydrodynamics, \texttt{ECHO}, has been very easily ported
to GPU-based systems (in particular on NVIDIA Volta and Ampere GPUs).

Our aim was twofold, not only to accelerate the code, but also to do this 
following the so-called standard language parallelism paradigm, meaning that
the original code must be fully preserved, by relying on just compiling options to recognise
the most modern language structures (and on the addition of few directives, 
simply ignored by the compiler when acceleration is not needed).
The version presented here is thus basically the original one, and now the very same
code can run indifferently on a laptop, on multiple CPU-based cores, or on GPU
accelerated devices.

Modern Fortran (the language in which \texttt{ECHO} was already written)
provided the easiest solution to achieve our goals. The use of standard ISO Fortran structures, 
such as \texttt{do concurrent} (DC) loops and \texttt{pure} subroutines and functions, 
allows one to accelerate all the computationally intensive loops (when the order
of execution is not an issue, true for many codes for fluid dynamics or MHD).
Directives are provided by the widely used OpenACC API, but these are actually not
necessary for data movement or loop acceleration, while we just use them for offloading the 
subroutines and functions invoked inside the accelerated loops (even this task would be hopefully 
doable by the compiler itself in the future). 
Full acceleration of DC loops and use of the CUDA Unified Memory address space (necessary
for allocating arrays on the device and providing automatic data management between the host
and the device) are simply activated by specific flags of the \texttt{nvfortran} compiler by NVIDIA,
while message passing is ensured by the standard MPI library for domain decomposition 
and communication among tasks.

As far as performance on CINECA supercomputers is concerned, the code is about 16 times faster 
when using the 4 GPUs of one node on LEONARDO (with the NVIDIA compiler)
compared to runs on the best performing CPU-based machine 
(GALILEO100 with 32 cores, $\times 4$ hyper-threading and Intel compiler with
aggressive optimization), at the resolution of $128^3$.
For larger sizes, and when LEONARDO is also employed for CPU multicore runs, the improvement
is even higher, up to $\simeq 30$ for a $512^3$ run.
The highest peak performance using the 4 GPUs of a single node of LEONARDO is 
$2.2 \times 10^8$ cells updated each iteration per second,
reached at both $512^3$ and $640^3$ resolution, for the second-order version of the code.
These numbers are quite important, it means that a CPU-based simulation may require 
from 15 to 30 nodes to achieve the same result, in the same time, obtained by using just
a single node employing the 4 GPUs.

Parallel scaling is very good within a single node, otherwise efficiency drops, unless the
ratio of the time spent in computation over communication is large, and this
is basically always true when GPUs are needed.
The simplest version of the code, where MPI calls are at a minimum, has a very good
strong scaling for $512^3$ cells, and a nearly ideal one for $1024^3$ cells.
The weak scaling test, using $512^3$ cells per node, shows an efficiency loss of just about 
$\sim 10 - 15\%$ in the range from 1 to 256 nodes (up to 1024 GPUs, corresponding to a 
maximum size of $4096 \times 4096 \times 2048  \simeq 34.4$ billion cells).
These performances worsen if full high order and the UCT method are both enforced,
given the numerous and the non GPU-aware communications between tasks (that is among GPUs).
To improve both strong and weak scaling also in the most complete case, we plan to enlarge the size 
of the halo regions of ghost cells to be exchanged among MPI tasks, at the price of increasing 
computations, so that message passing is performed just once per timestep in a single call, 
for all variables. Parallel efficiency is expected to raise above $90\%$, 
however we prefer to leave this task as future work.

The code will be soon applied to the study of 3D (special) relativistic MHD turbulence,
here we have just provided an example in 2D, showing simulations 
of decaying Alfvénic turbulence for a strongly magnetized and relativistically hot plasma. 
Small-scale vortices and current sheets are ubiquitous, and two full decades of inertial
range in the (isotropized) spectra, with a clear Kolmogorov $-5/3$ slope, are obtained
(for a resolution corresponding to $4096^2$ cells).
Numerical dissipation only affects the finest scales and appears to be well behaved, 
given that no accumulation of energy at the smallest scales is visible.
Since the wall clock time taken by an Alfvén wave to cross a 3D domain is estimated
to be of about an hour for a resolution of $4096^3$ using a full allocation of 256 nodes on LEONARDO,
a 3D simulation of decaying turbulence at the same resolution is expected to last just a few hours.

Moreover, being the code already capable of working in any curved metric of
general relativity, we are also ready to apply the new version of  \texttt{ECHO}
to the study of plasma surrounding compact objects, for example disk accretion
onto static and rotating black holes, extending to 3D our previous works for axisymmetric flows
\cite{Tomei:2020,DelZanna:2022}.

Concluding, we deem that basically any code for fluid dynamics or MHD written in modern
Fortran could be easily ported to GPUs (and heterogeneous architectures) with
a minimal effort by following our guidelines, hence we hope that our example may
be useful to many other scientists and engineers.
Maybe surprisingly, it really seems that the Fortran programming language, originally designed in the 
fifties simply for \textit{FORmula TRANslation} algorithms, can still be a leading language for 
science and engineering software applications, even nowadays in the pre-exascale era.
We also strongly believe that the standard language parallelism paradigm should be the 
path to follow in the future of HPC.

\vspace{6pt}
\authorcontributions{Conceptualization, L.D.Z., S.L., L.S., M.B., and E.P.; data curation, 
L.D.Z. and L.S.; supervision, L.D.Z. and S.L.; writing, L.D.Z. All authors have read and 
agreed to the published version of the manuscript.}

\funding{LDZ and SL acknowledge support from the ICSC---Centro 
Nazionale  di Ricerca in 
High-Performance Computing; Big Data and Quantum Computing, funded by European 
Union---NextGenerationEU.
MB acknowledges support by the European Union’s Horizon Europe research
and innovation program under the Marie Skłodowska-Curie grant agreement No. 
101064953.}

\dataavailability{Data are contained within the article.}

\acknowledgments{This work would not have been possible without the precious guidance by S. Deldon, 
our NVIDIA tutor
during the June 2022 (online) hackathon at CINECA. We also thank F. Spiga, S. Cielo, A. Mignone, 
and M. Rossazza for fruitful discussions, and the three anonymous referees for comments 
and suggestions, especially for asking us to extend the weak scaling test.
Numerical calculations have been made possible through a CINECA ISCRA-C grant (PI L. Del Zanna)
and the CINECA-INFN agreement, providing access to resources on MARCONI100, GALILEO100, 
and especially LEONARDO, at CINECA.}

\conflictsofinterest{The authors declare no conflict of interest.}


\end{paracol}








\reftitle{References}



\end{document}